\documentclass[12pt]{article}
\usepackage[dvips]{graphicx}
\usepackage{epsfig}
\usepackage{amsmath}
\usepackage{amssymb}
\usepackage{amsfonts}

\input{tcilatex}
\begin{document}

\title{Differential equation for Local Magnetization in the Boundary Ising
Model}
\author{Oleg Miroshnichenko\thanks{%
E-mail: oleg2@inbox.ru} \\
%EndAName
\\
{\normalsize \ }\textit{Bogolyubov Institute for Theoretical Physics, Kiev
03143, Ukraine}}
\date{}
\maketitle

\begin{abstract}
We show that the local magnetization in the massive boundary Ising model on
the half-plane with boundary magnetic field satisfies second order linear
differential equation whose coefficients are expressed through Painleve
function of the III kind.
\end{abstract}

\section{Introduction}

In the work \cite{FZ1} a very simple and elegant derivation of the famous
Painleve equations for the spin-spin correlation function in the scaling
Ising model with zero magnetic field was given. The approach used in that
work was also applied in \cite{FZ2} to derive finite volume form factors of
spin field in the Ising theory and in \cite{DF} to derive the differential
equation for spin-spin correlation functions in the Ising theory on a
pseudosphere. Here we apply this approach to derive differential equation
for local magnetization (i.~e. one-point correlation function of spin field)
in the boundary Ising model on the half-plane with boundary magnetic field.
It turns out to be a second order linear differential equation whose
coefficients are expressed through Painleve function of the III kind.
Supplied with appropriate boundary condition it uniquely defines the local
magnetization as a function of the distance to the boundary. Besides being
interesting in itself, such a representation for local magnetization may be
more convenient for numerical calculation in comparison with conventional
form factor expansion, especially in the short distance region.

As is well known \cite{C}, there are two essentially different types of
conformal boundary conditions (b.~c.) in conformal Ising field theory. The
so called "free" b.~c. corresponds to the universality class represented by
the lattice Ising model with unrestricted spins on the boundary. The so
called "fixed" b.~c. corresponds to the universality class represented by
the lattice Ising model with boundary spins all fixed in the same direction
("$+$" or "$-$", so there is more precisely two different "fixed" b.~c.).
All this b.~c. correspond to the fixed points of the boundary
renormalization group flow. The most general local b.~c. in the Ising field
theory is the "free" b.~c. perturbed by the boundary spin operator (which is
the only non-trivial relevant boundary operator in the case of "free"
b.~c.). This b.~c. corresponds to the renormalization group flow from "free"
b.~c. towards one of the "fixed" b.~c. \cite{AL}. More generally, one may
consider conformal Ising field theory with "free" b.~c., perturbed by both
boundary spin operator and bulk thermal operator \cite{GhZ}. This theory
describes the continuum limit in the vicinity of the critical point of the
lattice Ising model with zero magnetic field in the bulk and with boundary
magnetic field being suitably rescaled. Let us briefly list known results
about the local magnetisation in this theory defined on the half-plane.

The form factor expansion for local magnetization $\bar{\sigma}\left(
t\right) $ was written down in \cite{KLCM} using exact expression for
boundary state obtained in \cite{GhZ}:%
\begin{equation}
\bar{\sigma}\left( t\right) =\sigma _{0}\exp \left(
\dsum\limits_{k=1}^{\infty }\frac{1}{k}f_{k}\right)  \label{1a}
\end{equation}%
\begin{equation}
f_{k}=-\frac{1}{\pi ^{2}}\int\limits_{0}^{\infty }du_{1}\ldots
\int\limits_{0}^{\infty }du_{k}\dprod\limits_{l=1}^{k}\frac{\func{ch}u_{l}-1%
}{\func{ch}u_{l}+\func{ch}u_{l+1}}\left( \frac{\func{ch}u_{l}+1-\lambda }{%
\func{ch}u_{l}-1+\lambda }\right) e^{-t\func{ch}u_{l}}  \label{1b}
\end{equation}%
\begin{equation}
t=2m\text{y,\qquad }\lambda =\frac{4\pi h^{2}}{m}\text{,\qquad }\sigma
_{0}=2^{\frac{1}{12}}e^{-\frac{1}{8}}A^{\frac{3}{2}}m^{\frac{1}{8}}
\label{1c}
\end{equation}%
where $m\sim T-T_{c}$ is the mass of a particle, $h$ - scaling boundary
magnetic field, y - distance from the boundary, $\sigma _{0}$ -
magnetization on the infinite plane, $A=1.28243\ldots $ is Glaisher's
constant. Here and later on we always consider the low temperature phase $%
T<T_{c}$, unless it is specially pointed out. It is also implied conformal
normalization of spin field:%
\begin{equation}
\left\vert x-x^{\prime }\right\vert ^{\frac{1}{4}}\left\langle \sigma \left(
x\right) \sigma \left( x^{\prime }\right) \right\rangle \rightarrow 1\text{%
,\quad as }x\rightarrow x^{\prime }
\end{equation}%
With this normalization $\bar{\sigma}\left( t,\lambda \right) \rightarrow
\sigma _{0}$ as $t\rightarrow \infty $. The expansion (\ref{1a})-(\ref{1c})
was first obtained in \cite{B2} from lattice model calculations. It was also
shown in \cite{B1},\cite{B2} that in the cases of "free" ($h=0$) and "fixed"
($h\rightarrow \pm \infty $) b.~c. local magnetization can be expressed
through Painleve function of the III kind:%
\begin{equation}
\bar{\sigma}_{free}\left( t\right) =\sigma _{0}\exp \left\{ \frac{1}{4}%
\varphi \left( t\right) +\frac{1}{4}\int_{t}^{\infty }\left[ e^{-\varphi
\left( r\right) }-1+\frac{r}{2}\left( \func{sh}^{2}\varphi \left( r\right)
-\left( \varphi ^{\prime }\left( r\right) \right) ^{2}\right) \right]
dr\right\}  \label{3}
\end{equation}%
\begin{equation}
\bar{\sigma}_{fixed}\left( t\right) =\sigma _{0}\exp \left\{ -\frac{1}{4}%
\varphi \left( t\right) +\frac{1}{4}\int_{t}^{\infty }\left[ 1-e^{\varphi
\left( r\right) }+\frac{r}{2}\left( \func{sh}^{2}\varphi \left( r\right)
-\left( \varphi ^{\prime }\left( r\right) \right) ^{2}\right) \right]
dr\right\}  \label{4}
\end{equation}%
where $\varphi \left( r\right) $ is the solution of radial sinh-Gordon
equation:%
\begin{equation}
\varphi ^{\prime \prime }+\frac{1}{r}\varphi ^{\prime }=\frac{1}{2}\func{sh}%
2\varphi  \label{5}
\end{equation}%
satisfying asymptotic conditions:%
\begin{equation}
\varphi \left( r\right) =-\ln \left( -\frac{1}{2}r\Omega \right) +\emph{O}%
\left( r^{4}\Omega ^{2}\right) \text{, \quad as }r\rightarrow 0\text{,\quad }%
\Omega =\ln \left( \frac{e^{\gamma }}{8}r\right)  \label{6}
\end{equation}

\begin{equation}
\varphi \left( r\right) =\frac{2}{\pi }K_{0}\left( r\right) +\emph{O}\left(
e^{-3r}\right) \text{,\quad as }r\rightarrow \infty  \label{7}
\end{equation}%
where $\gamma $ is the Euler's constant, $K_{0}\left( x\right) $ is the
modified Bessel function of zeroth order. As is known, $\varphi \left(
x\right) $ is related to Painleve function of the III kind $\eta \left(
x\right) $ as $\eta \left( x\right) =e^{-\varphi \left( 2x\right) }$. More
about this function see \cite{MCWTB},\cite{MCTW}.

In the case when bulk is critical ($m=0$) it was shown in \cite{ChZ} that:%
\begin{equation}
\bar{\sigma}\left( \text{y}\right) =h2^{\frac{5}{4}}\pi ^{\frac{1}{2}}\left(
2\text{y}\right) ^{\frac{3}{8}}\Psi \left( 1/2,1,8\pi h^{2}\text{y}\right)
\label{8}
\end{equation}%
where 
\begin{equation}
\Psi \left( a,c,x\right) =\frac{1}{\Gamma \left( a\right) }\int_{0}^{\infty
}e^{-xt}t^{a-1}\left( 1+t\right) ^{c-a-1}dt  \label{9}
\end{equation}%
is a solution of degenerate hypergeometric equation.

The qualitative behavior of $\bar{\sigma}\left( t,\lambda \right) $ is well
understood \cite{KLCM}. On the whole interval $\left( 0,\infty \right) $ $%
\bar{\sigma}_{free}\left( t\right) $ monotonically increases and $\bar{\sigma%
}_{fixed}\left( t\right) $ monotonically decreases, both approaching $\sigma
_{0}$ as $t\rightarrow \infty $. For small $t$:\footnote{%
Note that comparing the coefficient in the short distance asymptotic of $%
\sigma _{fixed}\left( t\right) $, that follows from (\ref{4}), with the
result $\sigma _{fixed}\left( \text{y}\right) =2^{\frac{1}{4}}\left( 2\text{y%
}\right) ^{-\frac{1}{8}}$ (for $m=0$) of \cite{CL}, one obtains the identity 
$\int_{0}^{\infty }\left( 1-e^{-\varphi \left( r\right) }\right) dr=\ln 2$}%
\begin{equation}
\bar{\sigma}_{free}\left( t\right) \sim t^{\frac{3}{8}}  \label{10}
\end{equation}%
\begin{equation}
\bar{\sigma}_{fixed}\left( t\right) \sim t^{-\frac{1}{8}}  \label{11}
\end{equation}%
For $0<\lambda <2$ $\bar{\sigma}\left( t,\lambda \right) $ remains
monotonically increasing. Its values near the boundary are somewhat enhanced
by the presence of boundary magnetic field, the leading term of its short
distance asymptotic become dressed by logarithm:%
\begin{equation}
\bar{\sigma}\left( t\right) \sim t^{\frac{3}{8}}\ln t  \label{12}
\end{equation}%
For $\lambda >2$ it possesses a maximum in some point. As $\lambda
\rightarrow \infty $ this maximum turns into a very sharp peak located in
the region $t\sim \lambda ^{-1}$ near the boundary, its shape being
described by (\ref{8}), (\ref{9}). For $t\ll \lambda ^{-1}$ $\bar{\sigma}%
\left( t,\lambda \right) $ behaves as (\ref{12}), while for $t\gg \lambda
^{-1}$ its behavior coincides with one under "fixed" b.~c. (\ref{4}). This
dependence reflects the renormalization group cross-over between "free" and
"fixed" b.~c.

The main result of this paper is that for arbitrary $\lambda $:%
\begin{equation}
\bar{\sigma}\left( t,\lambda \right) =u\left( t,\lambda \right) \bar{\sigma}%
_{free}\left( t\right)  \label{15}
\end{equation}%
where $\bar{\sigma}_{free}\left( t\right) $ is given by (\ref{3}), and $%
u\left( t,\lambda \right) $ is the solution of differential equation:%
\begin{equation}
u^{\prime \prime }-\left( \varphi ^{\prime }-\func{ch}\varphi +\lambda
\right) u^{\prime }+\frac{1}{2}\lambda \left( \varphi ^{\prime }-\func{ch}%
\varphi +1\right) u=0  \label{16}
\end{equation}%
satisfying asymptotic condition:%
\begin{equation}
u\left( t\right) =1+\emph{O}\left( t^{-\frac{1}{2}}e^{-t}\right) \text{%
,\quad as }t\rightarrow \infty  \label{17}
\end{equation}%
(Here $\varphi \left( t\right) $ is the same function as in (\ref{3}), (\ref%
{4}) and the strokes stand for detivatives with respect to $t$.)

Let us make some remarks on the equation (\ref{16}). One can see that when $%
\lambda =0$, the only solution of (\ref{16}) satisfying (\ref{17}) is $%
u\left( t\right) =1$. When $\lambda \rightarrow \infty $ (\ref{16}) turns
into a first order differential equation which upon integrating and fixing
integration constant with the help of (\ref{17}) yields (\ref{4}). In the
massless limit ($t\rightarrow 0$, $\lambda \rightarrow \infty $, $t\lambda $
kept fixed) (\ref{16}) turns into a degenerate hypergeometric equation. Its
solution can be fixed by "sewing" its asymptotic as $t\lambda \rightarrow
\infty $ with asymptotic of (\ref{4}) as $t\rightarrow 0$, and this yields (%
\ref{8}), (\ref{9}). The fact that in massless limit we reproduce the result
of \cite{ChZ} is not very surprising because the approach we used to derive (%
\ref{15}), (\ref{16}) is a generalization of one used in \cite{ChZ}.
Concerning the relation between the form factor expansion (\ref{1a}), (\ref%
{1b}), (\ref{1c}) and our result (\ref{15}), (\ref{16}) we just note that it
seems to be very difficult to show directly that (\ref{1a}), (\ref{1b})
satisfy (\ref{15}), (\ref{16}). In any case, it is beyond the analytic
abilities of the author.

Being second order linear differential equation, (\ref{16}) possesses two
linearly independent solutions. Their asymptotics as $t\rightarrow \infty $
are $u_{1}\left( t\right) \sim 1$ and $u_{2}\left( t\right) \sim e^{\left(
\lambda -1\right) t}$. Hence, for $\lambda >1$ the condition that $u\left(
t\right) \rightarrow 1$ as $t\rightarrow \infty $ is sufficient to fix the
solution uniquely. For $\lambda \leq 1$ more strict condition (\ref{17}) is
required, which follows from form factor expansion (\ref{1a}), (\ref{1b}).
Another linearly independent solution in this case also has physical
meaning. As explained in \cite{GhZ}, for $\lambda <1$ there exists
metastable state characterized by asymptotic behavior $\bar{\sigma}\left(
t\right) \rightarrow -\sigma _{0}$ as $t\rightarrow \infty $ and
corresponding to the boundary bound state in the hamiltonian picture with
"space" being half-line and "time" axis being parallel to the boundary. The
local magnetization $\bar{\sigma}_{1}\left( t,\lambda \right) $ in this
state can be obtained from $\bar{\sigma}\left( t,\lambda \right) $ by
analytic continuation $h\rightarrow -h$. Clearly, it is also a solution of (%
\ref{16}). As it was shown in \cite{SchE} its asymptotic as $t\rightarrow
\infty $ is:%
\begin{equation}
\bar{\sigma}_{1}\left( t,\lambda \right) =-\sigma _{0}+\sigma _{0}\left( 
\frac{\lambda }{2-\lambda }\right) ^{\frac{1}{2}}e^{-\left( 1-\lambda
\right) t}+\frac{\sigma _{0}}{4\sqrt{2\pi }}\left( \frac{2}{\lambda }%
-1\right) t^{-\frac{3}{2}}e^{-t}+\emph{o}\left( t^{-\frac{3}{2}}e^{-t}\right)
\label{18}
\end{equation}%
The presence of exponential term $\sim e^{-\left( 1-\lambda \right) t}$ in (%
\ref{18}) agrees with (\ref{16}).

In the rest of the paper we present the details of our derivation of (\ref%
{15}), (\ref{16}).

\section{Ising field theory in the bulk}

In this section we briefly recall some well known facts \cite{FZ1} about the
structure of the Ising field theory in the bulk needed for further
computations. As is known the Ising field theory in zero magnetic field is
equivalent to the free Majorana fermion theory with euclidean action:%
\begin{equation}
S=\frac{1}{2\pi }\int \left( \psi \bar{\partial}\psi +\bar{\psi}\partial 
\bar{\psi}-im\bar{\psi}\psi \right) d^{2}x  \label{1-1}
\end{equation}%
Here we have assumed that the theory is defined on an infinite plane $%
%TCIMACRO{\U{211d} }%
%BeginExpansion
\mathbb{R}
%EndExpansion
^{2}$, whose points $x$ are labelled by cartesian coordinates $\left( \text{x%
},\,\text{y}\right) =\left( \text{x}\left( x\right) ,\,\text{y}\left(
x\right) \right) $, and $d^{2}x\equiv d$x$\,d$y. Complex coordinates are
defined as $z\left( x\right) =\,$x$\,+\,i$y, $\bar{z}\left( x\right) =\,$x $%
+\,i$y, and the derivatives $\partial ,\bar{\partial}$ in (\ref{1-1}) stand
for $\partial _{z}=\frac{1}{2}\left( \partial _{\text{x}}-i\partial _{\text{y%
}}\right) $ and $\partial _{\bar{z}}=\frac{1}{2}\left( \partial _{\text{x}%
}+i\partial _{\text{y}}\right) $ respectively. The ciral components $\psi ,$ 
$\bar{\psi}$ of fermi field satisfy Dirac's equations:%
\begin{equation}
\bar{\partial}\psi =-\frac{im}{2}\bar{\psi}\text{,\qquad }\partial \bar{\psi}%
=\frac{im}{2}\psi  \label{1-2}
\end{equation}%
Their normalization in the action (\ref{1-1}) corresponds to the following
short-distance limit of the operator products%
\begin{equation}
z\psi \left( x\right) \psi \left( 0\right) \rightarrow 1,\qquad \bar{z}\bar{%
\psi}\left( x\right) \bar{\psi}\left( 0\right) \rightarrow 1,\qquad \text{as 
}x\rightarrow 0  \label{1-2a}
\end{equation}

The order $\sigma \left( x\right) $ and disorder $\mu \left( x\right) $
fields are semi-local with respect to the fermi fields; the products%
\begin{equation}
\psi \left( x\right) \sigma \left( 0\right) ,\qquad \psi \left( x\right) \mu
\left( 0\right) ,\qquad \bar{\psi}\left( x\right) \sigma \left( 0\right)
,\qquad \bar{\psi}\left( x\right) \mu \left( 0\right)  \label{1-3}
\end{equation}%
acquire a minus sign when the point $x$ is taken around zero point. The
fields $\psi \left( x\right) $ and $\bar{\psi}\left( x\right) $ in the
products (\ref{1-3}) can be expanded in the complete set of solutions of
Dirac's equations (\ref{1-2}) having this monodromy property:%
\begin{equation}
\left( 
\begin{array}{c}
\psi \left( x\right) \\ 
\bar{\psi}\left( x\right)%
\end{array}%
\right) =\sum_{n\in 
%TCIMACRO{\U{2124} }%
%BeginExpansion
\mathbb{Z}
%EndExpansion
}a_{n}\left( 
\begin{array}{c}
u_{-n}\left( x\right) \\ 
\bar{u}_{-n}\left( x\right)%
\end{array}%
\right) +\bar{a}_{n}\left( 
\begin{array}{c}
v_{-n}\left( x\right) \\ 
\bar{v}_{-n}\left( x\right)%
\end{array}%
\right)  \label{1-4}
\end{equation}%
where%
\begin{equation}
\left( 
\begin{array}{c}
u_{n}\left( x\right) \\ 
\bar{u}_{n}\left( x\right)%
\end{array}%
\right) =\left( \frac{m}{2}\right) ^{\frac{1}{2}-n}\Gamma \left( n+\frac{1}{2%
}\right) \left( 
\begin{array}{c}
e^{i\left( n-\frac{1}{2}\right) \theta }I_{n-\frac{1}{2}}\left( mr\right) \\ 
-ie^{i\left( n+\frac{1}{2}\right) \theta }I_{n+\frac{1}{2}}\left( mr\right)%
\end{array}%
\right)  \label{1-5}
\end{equation}%
\begin{equation}
\left( 
\begin{array}{c}
v_{n}\left( x\right) \\ 
\bar{v}_{n}\left( x\right)%
\end{array}%
\right) =\left( \frac{m}{2}\right) ^{\frac{1}{2}-n}\Gamma \left( n+\frac{1}{2%
}\right) \left( 
\begin{array}{c}
ie^{-i\left( n+\frac{1}{2}\right) \theta }I_{n+\frac{1}{2}}\left( mr\right)
\\ 
e^{-i\left( n-\frac{1}{2}\right) \theta }I_{n-\frac{1}{2}}\left( mr\right)%
\end{array}%
\right)  \label{1-6}
\end{equation}%
(here $r$, $\theta $ are polar coordinates, i.~e. $z=re^{i\theta }$, $\bar{z}%
=re^{-i\theta }$ and $I_{\nu }$ are modified Bessel functions). The
coefficients $a_{n}$, $\bar{a}_{n}$ in (\ref{1-4}) are understood as
operators acting on the space of fields.

It can be easily shown that for any two solutions $\Psi _{1}=\left( \psi
_{1}\left( x\right) ,\bar{\psi}_{1}\left( x\right) \right) $, $\Psi
_{2}=\left( \psi _{2}\left( x\right) ,\bar{\psi}_{2}\left( x\right) \right) $
of Dirac's equations which change sign after the point $x$ is taken around
zero the integral%
\begin{equation}
\left( \Psi _{1},\Psi _{2}\right) =\frac{1}{2\pi i}\oint\limits_{C_{0}}\psi
_{1}\left( x\right) \psi _{2}\left( x\right) dz-\bar{\psi}_{1}\left(
x\right) \bar{\psi}_{2}\left( x\right) d\bar{z}  \label{1-6-1}
\end{equation}%
over a contour $C_{0}$ encircling zero (in counter-clockwise direction) does
not change under continuous deformation of $C_{0}$ and therefore defines a
bilinear form on the space of such solutions. The solutions $U_{n}=\left(
u_{n},\bar{u}_{n}\right) $ and $V_{n}=\left( v_{n},\bar{v}_{n}\right) $
satisfy the following orthogonality properties with respect to this bilinear
form:%
\begin{equation}
\left( U_{n},U_{m}\right) =\delta _{n+m,0}\text{ ,\qquad }\left(
V_{n},V_{m}\right) =\delta _{n+m,0}\text{ ,}\qquad \left( U_{n},V_{m}\right)
=0  \label{1-6-2}
\end{equation}%
Let us also write down the following differentiation formulas which are
useful in computations with $U_{n}$ and $V_{n}$:%
\begin{eqnarray}
\partial U_{n} &=&\left( n-\frac{1}{2}\right) U_{n-1}\text{ ,\qquad }\bar{%
\partial}U_{n}=\frac{m^{2}}{2\left( 2n+1\right) }U_{n+1}  \notag \\[0.05in]
\partial V_{n} &=&\frac{m^{2}}{2\left( 2n+1\right) }V_{n+1}\text{ ,\qquad }%
\bar{\partial}V_{n}=\left( n-\frac{1}{2}\right) V_{n-1}  \label{1-6-3}
\end{eqnarray}%
(here we denote $\partial U_{n}\equiv \left( \partial u_{n}\left( x\right)
,\partial \bar{u}_{n}\left( x\right) \right) $, etc.).

Using relations (\ref{1-6-2}) one can express operators $a_{n}$, $\bar{a}%
_{n} $ in terms of contour integrals:%
\begin{equation}
a_{n}=\frac{1}{2\pi i}\oint\limits_{C_{0}}u_{n}\left( x\right) \psi \left(
x\right) dz-\bar{u}_{n}\left( x\right) \bar{\psi}\left( x\right) d\bar{z}
\label{1-6-4}
\end{equation}%
\begin{equation}
\bar{a}_{n}=\frac{1}{2\pi i}\oint\limits_{C_{0}}v_{n}\left( x\right) \psi
\left( x\right) dz-\bar{v}_{n}\left( x\right) \bar{\psi}\left( x\right) d%
\bar{z}  \label{1-6-5}
\end{equation}%
This representation can be used to show that they satisfy canonical
commutation relations:%
\begin{equation}
\left\{ a_{n},a_{m}\right\} =\delta _{n+m,0}\text{ ,}\qquad \left\{ \bar{a}%
_{n},\bar{a}_{m}\right\} =\delta _{n+m,0}\text{ ,}\qquad \left\{ a_{n},\bar{a%
}_{m}\right\} =0  \label{1-7}
\end{equation}%
The fields $\sigma $ and $\mu $ are "primary" with respect to the algebra (%
\ref{1-7}), i.~e. they satisfy relations:%
\begin{equation}
a_{n}\sigma =0\text{ ,}\qquad \bar{a}_{n}\sigma =0\text{ ,\qquad }a_{n}\mu =0%
\text{ ,\qquad }\bar{a}_{n}\mu =0  \label{1-8}
\end{equation}%
for $n>0$, as well as%
\begin{eqnarray}
a_{0}\sigma &=&\frac{\omega }{\sqrt{2}}\mu \text{ ,}\qquad a_{0}\mu =\frac{%
\bar{\omega}}{\sqrt{2}}\sigma  \notag \\[0.05in]
\bar{a}_{0}\sigma &=&\frac{\bar{\omega}}{\sqrt{2}}\mu \text{ ,}\qquad \bar{a}%
_{0}\mu =\frac{\omega }{\sqrt{2}}\sigma  \label{1-9}
\end{eqnarray}%
where $\omega =e^{i\pi /4}$ and $\bar{\omega}=e^{-i\pi /4}$. These equations
define the fields $\sigma $ and $\mu $ up to normalization. In what follows
we will assume conformal normalization of fields $\sigma $ and $\mu $:%
\begin{equation}
\left\vert x\right\vert ^{\frac{1}{4}}\sigma \left( x\right) \sigma \left(
0\right) \rightarrow 1\text{ ,\qquad }\left\vert x\right\vert ^{\frac{1}{4}%
}\mu \left( x\right) \mu \left( 0\right) \rightarrow 1\text{ ,\qquad as }%
x\rightarrow 0  \label{1-11}
\end{equation}%
As it is shown in \cite{FZ1}, first and second order descendants of $\sigma $
and $\mu $ with respect to the algebra $a_{n}$, $\bar{a}_{n}$ are expressed
in terms of coordinate derivatives of $\sigma $ and $\mu $:%
\begin{eqnarray}
a_{-1}\sigma &=&\frac{\omega }{\sqrt{2}}4\partial \mu \text{ ,}\qquad
a_{-1}\mu =\frac{\bar{\omega}}{\sqrt{2}}4\partial \sigma  \notag \\[0.05in]
\bar{a}_{-1}\sigma &=&\frac{\bar{\omega}}{\sqrt{2}}4\bar{\partial}\mu \text{
,}\qquad \bar{a}_{-1}\mu =\frac{\omega }{\sqrt{2}}4\bar{\partial}\sigma
\label{1-12}
\end{eqnarray}%
\begin{eqnarray}
a_{-2}\sigma &=&\frac{\omega }{\sqrt{2}}\frac{8}{3}\partial ^{2}\mu \text{ ,}%
\qquad a_{-2}\mu =\frac{\bar{\omega}}{\sqrt{2}}\frac{8}{3}\partial ^{2}\sigma
\notag \\[0.05in]
\bar{a}_{-2}\sigma &=&\frac{\bar{\omega}}{\sqrt{2}}\frac{8}{3}\bar{\partial}%
^{2}\mu \text{ ,}\qquad \bar{a}_{-2}\mu =\frac{\omega }{\sqrt{2}}\frac{8}{3}%
\bar{\partial}^{2}\sigma  \label{1-13}
\end{eqnarray}%
This observation is very important for the method of \cite{FZ1} to work.

The Majorana theory (\ref{1-1}) corresponds to both high and low temperature
phases of the Ising model in the vicinity of its critical point $T_{c}$
depending of the choice of the sign of the mass parameter $m$ in (\ref{1-1}%
). Our definition in (\ref{1-9}) corresponds to the identification of the
case $m>0$ with the ordered phase $T<T_{c}$, while the case $m<0$ is
identified with the disordered phase $T>T_{c}$. From now on we will consider
the ordered phase $m>0$.

\section{"Free" and "fixed" boundary conditions}

In this section we consider the Ising field theory defined on the half-plane
y $>0$. As a warm-up exercise let us first rederive formulas (\ref{3}), (\ref%
{4}) for local magnetization in the cases of "free" and "fixed" b.~c.
Explicit expressions for boundary states for "free" and "fixed" b.~c. was
obtained in \cite{C}. It follows from this expressions that the fields $\psi 
$ and $\bar{\psi}$ satisfy the following b.~c.:%
\begin{eqnarray}
\left. \left( \psi -\bar{\psi}\right) \right\vert _{\text{y}=0} &=&0\qquad 
\text{(for "free" b.~c.)}  \label{2-1} \\[0.05in]
\left. \left( \psi +\bar{\psi}\right) \right\vert _{\text{y}=0} &=&0\qquad 
\text{(for "fixed" b.~c.)}  \label{2-2}
\end{eqnarray}%
Suppose that $\left( \chi \left( x\right) ,\bar{\chi}\left( x\right) \right) 
$ is a double-valued solution of Dirac's equations defined on the half-plane
y $>0$ (with punctured point $x_{0}$) such that it changes sign when $x$ is
taken around $x_{0}$, decays sufficiently fast as $\left\vert x\right\vert
\rightarrow \infty $, and satisfies b.~c.:%
\begin{eqnarray}
\left. \left( \chi -\bar{\chi}\right) \right\vert _{\text{y}=0} &=&0\qquad 
\text{(for "free" b.~c.)}  \label{2-3} \\[0.05in]
\left. \left( \chi +\bar{\chi}\right) \right\vert _{\text{y}=0} &=&0\qquad 
\text{(for "fixed" b.~c.)}  \label{2-4}
\end{eqnarray}%
Then the following identity holds:%
\begin{equation}
\left\langle {}\right. \left( {}\right. \oint\limits_{C_{x_{0}}}\chi \left(
x\right) \psi \left( x\right) dz-\bar{\chi}\left( x\right) \bar{\psi}\left(
x\right) d\bar{z}\left. {}\right) \mu \left( x_{0}\right) \left.
{}\right\rangle =0  \label{2-5}
\end{equation}%
where $C_{x_{0}}$ is a contour encircling the point $x_{0}$. This is because
the integral on the left-hand side of (\ref{2-5}) does not changes under
continuous deformations of the contour $C_{x_{0}}$ and therefore one can
deform it in such a way that it constitutes of two parts: $C_{\infty }$
which tends to infinity and $C_{b}$ which passes along the boundary. Then
the integral along $C_{\infty }$ is zero because $\chi \left( x\right) $ and 
$\bar{\chi}\left( x\right) $ decay at infinity and the integral along $C_{b}$
is zero due to (\ref{2-3}) or (\ref{2-4}). On the other hand one can shrink $%
C_{x_{0}}$ to a small circle around the point $x_{0}$ and express the
left-hand side of (\ref{2-5}) in terms of descendents of $\mu $ using the
operator product expansions of $\psi \left( x\right) $, $\bar{\psi}\left(
x\right) $ with $\mu \left( x_{0}\right) $. If the singularity of $\left(
\chi \left( x\right) ,\bar{\chi}\left( x\right) \right) $ at the point $%
x_{0} $ is not too strong only descendents of not higher than the second
order appear in this expression. Since they are expressed in terms of
coordinate derivatives of $\sigma $ this will lead to a differential
equation for local magnetization $\left\langle \sigma \left( x\right)
\right\rangle $. The only problem is to find a solution of Dirac's equations 
$(\chi \left( x\right) ,\bar{\chi}\left( x\right) )$ satisfying the above
conditions. For this purpose we will use the following trick. We will search
for this solution in the form of the linear combination:%
\begin{multline}
\left( 
\begin{array}{c}
\chi \left( x\right) \\ 
\bar{\chi}\left( x\right)%
\end{array}%
\right) =c_{1}\left( 
\begin{array}{c}
\left\langle \psi \left( x\right) \sigma \left( x_{0}\right) \mu \left(
Px_{0}\right) \right\rangle _{0} \\ 
\left\langle \bar{\psi}\left( x\right) \sigma \left( x_{0}\right) \mu \left(
Px_{0}\right) \right\rangle _{0}%
\end{array}%
\right) +c_{2}\partial \left( 
\begin{array}{c}
\left\langle \psi \left( x\right) \sigma \left( x_{0}\right) \mu \left(
Px_{0}\right) \right\rangle _{0} \\ 
\left\langle \bar{\psi}\left( x\right) \sigma \left( x_{0}\right) \mu \left(
Px_{0}\right) \right\rangle _{0}%
\end{array}%
\right) +  \label{2-6} \\[0.05in]
+c_{3}\bar{\partial}\left( 
\begin{array}{c}
\left\langle \psi \left( x\right) \sigma \left( x_{0}\right) \mu \left(
Px_{0}\right) \right\rangle _{0} \\ 
\left\langle \bar{\psi}\left( x\right) \sigma \left( x_{0}\right) \mu \left(
Px_{0}\right) \right\rangle _{0}%
\end{array}%
\right) +c_{4}\left( 
\begin{array}{c}
\left\langle \psi \left( x\right) \mu \left( x_{0}\right) \sigma \left(
Px_{0}\right) \right\rangle _{0} \\ 
\left\langle \bar{\psi}\left( x\right) \mu \left( x_{0}\right) \sigma \left(
Px_{0}\right) \right\rangle _{0}%
\end{array}%
\right) + \\[0.05in]
+c_{5}\partial \left( 
\begin{array}{c}
\left\langle \psi \left( x\right) \mu \left( x_{0}\right) \sigma \left(
Px_{0}\right) \right\rangle _{0} \\ 
\left\langle \bar{\psi}\left( x\right) \mu \left( x_{0}\right) \sigma \left(
Px_{0}\right) \right\rangle _{0}%
\end{array}%
\right) +c_{6}\bar{\partial}\left( 
\begin{array}{c}
\left\langle \psi \left( x\right) \mu \left( x_{0}\right) \sigma \left(
Px_{0}\right) \right\rangle _{0} \\ 
\left\langle \bar{\psi}\left( x\right) \mu \left( x_{0}\right) \sigma \left(
Px_{0}\right) \right\rangle _{0}%
\end{array}%
\right)
\end{multline}%
where $\left\langle \ldots \right\rangle _{0}$ denotes a correlation
function in the Ising field theory defined on the plane and $P$ denotes the
reflection in the line y $=0$ (i.~e. $P\left( \text{x, y}\right) =\left( 
\text{x, }-\text{y}\right) $). Obviously each term in (\ref{2-6}) is
non-zero and as a function of $x$ is a solution of Dirac's equations, change
sign when $x$ is taken around $x_{0}$ and decay at infinity. The
coefficients in this linear combination can be determined from the
requirement that it satisfies (\ref{2-3}) or (\ref{2-4}). Note that we do
not need to know the functions $\chi \left( x\right) $ and $\bar{\chi}\left(
x\right) $ in explicit form. What we really need is several terms of their
short-distance asymptotics as $x\rightarrow x_{0}$, but the latter can be
expressed in terms of the two-point functions $\left\langle \sigma \left(
x_{0}\right) \sigma \left( Px_{0}\right) \right\rangle _{0}\equiv G\left( 2m%
\text{y}_{0}\right) $ and $\left\langle \mu \left( x_{0}\right) \mu \left(
Px_{0}\right) \right\rangle _{0}\equiv \tilde{G}\left( 2m\text{y}_{0}\right) 
$. As is well known \cite{MCWTB} (see also \cite{FZ1}) this functions can be
expressed in terms of Painleve function of the III kind as follows:%
\begin{equation}
G\left( t\right) =\sigma _{0}\func{ch}\left( \frac{1}{2}\varphi \left(
t\right) \right) \exp \left[ \frac{1}{4}\int_{t}^{\infty }r\left( \func{sh}%
^{2}\varphi \left( r\right) -\left( \varphi ^{\prime }\left( r\right)
\right) ^{2}\right) dr\right]  \label{2-6-1}
\end{equation}%
\begin{equation}
\tilde{G}\left( t\right) =\sigma _{0}\func{sh}\left( \frac{1}{2}\varphi
\left( t\right) \right) \exp \left[ \frac{1}{4}\int_{t}^{\infty }r\left( 
\func{sh}^{2}\varphi \left( r\right) -\left( \varphi ^{\prime }\left(
r\right) \right) ^{2}\right) dr\right]  \label{2-6-2}
\end{equation}%
where $\varphi \left( t\right) $ is the same function as in (\ref{3}), (\ref%
{4}).

Under parity transformation $P$ fermi fields transform as:%
\begin{equation}
P\left( 
\begin{array}{c}
\psi \\ 
\bar{\psi}%
\end{array}%
\right) =\left( 
\begin{array}{c}
-i\bar{\psi} \\ 
i\psi%
\end{array}%
\right)  \label{2-7}
\end{equation}%
We have therefore the following identities which follows from the invariance
of correlation functions under parity transformation:%
\begin{eqnarray}
\left. \left\langle \bar{\psi}\left( x\right) \sigma \left( x_{0}\right) \mu
\left( Px_{0}\right) \right\rangle _{0}\right\vert _{\text{y}=0} &=&\left.
i\left\langle \psi \left( x\right) \mu \left( x_{0}\right) \sigma \left(
Px_{0}\right) \right\rangle _{0}\right\vert _{\text{y}=0}  \label{2-8} \\%
[0.05in]
\left. \left\langle \bar{\psi}\left( x\right) \mu \left( x_{0}\right) \sigma
\left( Px_{0}\right) \right\rangle _{0}\right\vert _{\text{y}=0} &=&\left.
i\left\langle \psi \left( x\right) \sigma \left( x_{0}\right) \mu \left(
Px_{0}\right) \right\rangle _{0}\right\vert _{\text{y}=0}  \label{2-9} \\%
[0.05in]
\left. \left\langle \bar{\partial}\bar{\psi}\left( x\right) \sigma \left(
x_{0}\right) \mu \left( Px_{0}\right) \right\rangle _{0}\right\vert _{\text{y%
}=0} &=&\left. i\left\langle \partial \psi \left( x\right) \mu \left(
x_{0}\right) \sigma \left( Px_{0}\right) \right\rangle _{0}\right\vert _{%
\text{y}=0}  \label{2-10} \\[0.05in]
\left. \left\langle \bar{\partial}\bar{\psi}\left( x\right) \mu \left(
x_{0}\right) \sigma \left( Px_{0}\right) \right\rangle _{0}\right\vert _{%
\text{y}=0} &=&\left. i\left\langle \partial \psi \left( x\right) \sigma
\left( x_{0}\right) \mu \left( Px_{0}\right) \right\rangle _{0}\right\vert _{%
\text{y}=0}  \label{2-11}
\end{eqnarray}%
It follows from this identities and Dirac's equations that from all
correlation functions that present in the expression (\ref{2-6}) only four
are linearly independent functions of $x$ on the line y $=0$ (for example $%
\left\langle \psi \left( x\right) \sigma \left( x_{0}\right) \mu \left(
Px_{0}\right) \right\rangle $, $\left\langle \psi \left( x\right) \mu \left(
x_{0}\right) \sigma \left( Px_{0}\right) \right\rangle $, $\left\langle
\partial \psi \left( x\right) \sigma \left( x_{0}\right) \mu \left(
Px_{0}\right) \right\rangle $, and $\left\langle \partial \psi \left(
x\right) \mu \left( x_{0}\right) \sigma \left( Px_{0}\right) \right\rangle $%
). Hence requiring that (\ref{2-6}) satisfy (\ref{2-3}) or (\ref{2-4}) one
obtains four linear constraints for six coefficients $c_{1},\ldots c_{6}$.
It turns out that they have non-zero solutions. One of the solutions
corresponds to the following linear combination (it does not matter what of
the solutions to choose):%
\begin{multline}
\left( 
\begin{array}{c}
\chi \left( x\right) \\ 
\bar{\chi}\left( x\right)%
\end{array}%
\right) =\frac{m}{2}\left( 
\begin{array}{c}
\left\langle \psi \left( x\right) \sigma \left( x_{0}\right) \mu \left(
Px_{0}\right) \right\rangle _{0} \\ 
\left\langle \bar{\psi}\left( x\right) \sigma \left( x_{0}\right) \mu \left(
Px_{0}\right) \right\rangle _{0}%
\end{array}%
\right) -i\partial \left( 
\begin{array}{c}
\left\langle \psi \left( x\right) \sigma \left( x_{0}\right) \mu \left(
Px_{0}\right) \right\rangle _{0} \\ 
\left\langle \bar{\psi}\left( x\right) \sigma \left( x_{0}\right) \mu \left(
Px_{0}\right) \right\rangle _{0}%
\end{array}%
\right) +  \label{2-12} \\[0.05in]
+i\frac{m}{2}\left( 
\begin{array}{c}
\left\langle \psi \left( x\right) \mu \left( x_{0}\right) \sigma \left(
Px_{0}\right) \right\rangle _{0} \\ 
\left\langle \bar{\psi}\left( x\right) \mu \left( x_{0}\right) \sigma \left(
Px_{0}\right) \right\rangle _{0}%
\end{array}%
\right) -\bar{\partial}\left( 
\begin{array}{c}
\left\langle \psi \left( x\right) \mu \left( x_{0}\right) \sigma \left(
Px_{0}\right) \right\rangle _{0} \\ 
\left\langle \bar{\psi}\left( x\right) \mu \left( x_{0}\right) \sigma \left(
Px_{0}\right) \right\rangle _{0}%
\end{array}%
\right) \text{,}
\end{multline}%
for "free" b.~c., and%
\begin{multline}
\left( 
\begin{array}{c}
\chi \left( x\right) \\ 
\bar{\chi}\left( x\right)%
\end{array}%
\right) =i\frac{m}{2}\left( 
\begin{array}{c}
\left\langle \psi \left( x\right) \sigma \left( x_{0}\right) \mu \left(
Px_{0}\right) \right\rangle _{0} \\ 
\left\langle \bar{\psi}\left( x\right) \sigma \left( x_{0}\right) \mu \left(
Px_{0}\right) \right\rangle _{0}%
\end{array}%
\right) -\partial \left( 
\begin{array}{c}
\left\langle \psi \left( x\right) \sigma \left( x_{0}\right) \mu \left(
Px_{0}\right) \right\rangle _{0} \\ 
\left\langle \bar{\psi}\left( x\right) \sigma \left( x_{0}\right) \mu \left(
Px_{0}\right) \right\rangle _{0}%
\end{array}%
\right) +  \label{2-13} \\[0.05in]
+\frac{m}{2}\left( 
\begin{array}{c}
\left\langle \psi \left( x\right) \mu \left( x_{0}\right) \sigma \left(
Px_{0}\right) \right\rangle _{0} \\ 
\left\langle \bar{\psi}\left( x\right) \mu \left( x_{0}\right) \sigma \left(
Px_{0}\right) \right\rangle _{0}%
\end{array}%
\right) -i\bar{\partial}\left( 
\begin{array}{c}
\left\langle \psi \left( x\right) \mu \left( x_{0}\right) \sigma \left(
Px_{0}\right) \right\rangle _{0} \\ 
\left\langle \bar{\psi}\left( x\right) \mu \left( x_{0}\right) \sigma \left(
Px_{0}\right) \right\rangle _{0}%
\end{array}%
\right) \text{,}
\end{multline}%
for "fixed" b.~c.

It is now straightforward but somewhat tedious exercise to substitute (\ref%
{2-12}) and (\ref{2-13}) in (\ref{2-5}) and evaluate the left-hand side. One
has to expand $\psi $ and $\bar{\psi}$ using (\ref{1-4}), than to evaluate
contour integrals using (\ref{1-6-2}), (\ref{1-6-3}) and than to act by the
operators $a_{n}$, $\bar{a}_{n}$ on $\sigma $ and $\mu $ using (\ref{1-8}), (%
\ref{1-9}) and (\ref{1-12}). Due to (\ref{1-8}) all terms with descendants
of higher than the first order vanish. Finally, taking into account that $%
\left\langle \sigma \left( x_{0}\right) \right\rangle \equiv \bar{\sigma}%
\left( 2m\text{y}_{0}\right) $ depends only on y$_{0}$ due to translation
invariance, one obtains the following differential equations:%
\begin{equation}
2\left( G-\tilde{G}\right) \bar{\sigma}_{free}^{\prime }-\left( G^{\prime }-%
\tilde{G}^{\prime }+\tilde{G}\right) \bar{\sigma}_{free}=0  \label{2-14}
\end{equation}%
\begin{equation}
2\left( G+\tilde{G}\right) \bar{\sigma}_{fixed}^{\prime }-\left( G^{\prime }+%
\tilde{G}^{\prime }+\tilde{G}\right) \bar{\sigma}_{fixed}=0  \label{2-15}
\end{equation}%
(the stroke denotes derivative with respect to $t=2m$y$_{0}$). Integrating
this equations, substituting (\ref{2-6-1}), (\ref{2-6-2}) and fixing
integration constants with the help of asymptotic condition $\bar{\sigma}%
\left( t\right) \rightarrow \sigma _{0}$ as $t\rightarrow \infty $ one
obtains (\ref{3}) and (\ref{4}).

Let us now consider the high temperature phase $T>T_{c}$. The differential
equations in this case can be obtained from (\ref{2-14}), (\ref{2-15}) by
substitution $m\rightarrow -m$, $G\rightleftarrows \tilde{G}$:%
\begin{equation}
2\left( G-\tilde{G}\right) \bar{\sigma}_{free}^{\prime }-\left( G^{\prime }-%
\tilde{G}^{\prime }+G\right) \bar{\sigma}_{free}=0  \label{2-16}
\end{equation}%
\begin{equation}
2\left( G+\tilde{G}\right) \bar{\sigma}_{fixed}^{\prime }-\left( G^{\prime }+%
\tilde{G}^{\prime }-G\right) \bar{\sigma}_{fixed}=0  \label{2-17}
\end{equation}%
The only solution of (\ref{2-16}) that does not grow exponentially as $%
t\rightarrow \infty $ is $\bar{\sigma}_{free}=0$, while from (\ref{2-17}) we
obtain:%
\begin{equation}
\bar{\sigma}_{fixed,\;T>T_{c}}=e^{-\frac{1}{2}t}\bar{\sigma}%
_{fixed,\;T<T_{c}}  \label{2-18}
\end{equation}%
in agreement with \cite{B2}. This confirms our identification of the case $%
m>0$ with the low temperature phase. Had we chosen the other choice, we
would obtain the exponentially growing solution for $\bar{\sigma}_{fixed}$
in the high temperature phase.

\section{Boundary magnetic field}

Let us now consider the general case of "free" b.~c. perturbed by boundary
spin operator $\sigma _{B}$. The latter is identified with degenerate
primary boundary field with dimension $\Delta =1/2$ \cite{C}. It can be
written in terms of fermion fields as follows \cite{GhZ}:%
\begin{equation}
\sigma _{B}\left( \text{x}\right) =ia\left( \text{x}\right) \left. \left(
\psi \left( x\right) +\bar{\psi}\left( x\right) \right) \right\vert _{\text{%
y=0}}  \label{3-1}
\end{equation}%
where $a\left( \text{x}\right) $ is additional fermionic degree of freedom
with two-point function%
\begin{equation}
\left\langle a\left( \text{x}\right) a\left( \text{x'}\right) \right\rangle
_{free}=\frac{1}{2}\,\text{sign}\left( \text{x}-\text{x'}\right)  \label{3-2}
\end{equation}%
The action of the theory has therefore the following form:%
\begin{multline}
S=\frac{1}{2\pi }\int\limits_{-\infty }^{\infty }d\text{x}%
\int\limits_{0}^{\infty }d\text{y}\left( \psi \bar{\partial}\psi +\bar{\psi}%
\partial \bar{\psi}-im\bar{\psi}\psi \right) +  \label{3-3} \\
+\int\limits_{-\infty }^{\infty }\left( -\frac{i}{4\pi }\left. \left( \psi 
\bar{\psi}\right) \right\vert _{\text{y}=0}+\frac{1}{2}a\partial _{\text{x}%
}a\right) d\text{x}+ih\int\limits_{-\infty }^{\infty }a\left( \text{x}%
\right) \left. \left( \psi +\bar{\psi}\right) \right\vert _{\text{y}=0}d%
\text{x}
\end{multline}%
It leads to the following b.~c. for fermion fields \cite{GhZ}:%
\begin{equation}
\frac{\partial }{\partial \text{x}}\left. \left( \psi -\bar{\psi}\right)
\right\vert _{\text{y}=0}=\left. -im\lambda \left( \psi +\bar{\psi}\right)
\right\vert _{\text{y}=0}  \label{3-4}
\end{equation}%
where $\lambda =4\pi h^{2}/m$. We can now proceed in the same way as in the
previous section but now instead of (\ref{2-3}) or (\ref{2-4}) we should
require the functions $\chi $ and $\bar{\chi}$ to satisfy the condition:

\begin{equation}
\frac{\partial }{\partial \text{x}}\left. \left( \chi -\bar{\chi}\right)
\right\vert _{\text{y}=0}=\left. -im\lambda \left( \chi +\bar{\chi}\right)
\right\vert _{\text{y}=0}  \label{3-5}
\end{equation}%
in order to write down the Ward identity (\ref{2-5}). It turns out that in
this case it is necessary to include also terms with second order
derivatives in the linear combination (\ref{2-6}) in order to satisfy (\ref%
{3-5}). As a result one obtains the following linear combination:%
\begin{multline}
\left( 
\begin{array}{c}
\chi \left( x\right) \\ 
\bar{\chi}\left( x\right)%
\end{array}%
\right) =i\left( \frac{m}{2}\right) ^{2}\left( 1-2\lambda \right) \left( 
\begin{array}{c}
\left\langle \psi \left( x\right) \sigma \left( x_{0}\right) \mu \left(
Px_{0}\right) \right\rangle _{0} \\ 
\left\langle \bar{\psi}\left( x\right) \sigma \left( x_{0}\right) \mu \left(
Px_{0}\right) \right\rangle _{0}%
\end{array}%
\right) -  \label{3-6} \\[0.05in]
-\frac{m}{2}\left( 1-2\lambda \right) \partial \left( 
\begin{array}{c}
\left\langle \psi \left( x\right) \sigma \left( x_{0}\right) \mu \left(
Px_{0}\right) \right\rangle _{0} \\ 
\left\langle \bar{\psi}\left( x\right) \sigma \left( x_{0}\right) \mu \left(
Px_{0}\right) \right\rangle _{0}%
\end{array}%
\right) -\frac{m}{2}\bar{\partial}\left( 
\begin{array}{c}
\left\langle \psi \left( x\right) \sigma \left( x_{0}\right) \mu \left(
Px_{0}\right) \right\rangle _{0} \\ 
\left\langle \bar{\psi}\left( x\right) \sigma \left( x_{0}\right) \mu \left(
Px_{0}\right) \right\rangle _{0}%
\end{array}%
\right) + \\[0.05in]
+i\partial ^{2}\left( 
\begin{array}{c}
\left\langle \psi \left( x\right) \sigma \left( x_{0}\right) \mu \left(
Px_{0}\right) \right\rangle _{0} \\ 
\left\langle \bar{\psi}\left( x\right) \sigma \left( x_{0}\right) \mu \left(
Px_{0}\right) \right\rangle _{0}%
\end{array}%
\right) +\left( \frac{m}{2}\right) ^{2}\left( 1-2\lambda \right) \left( 
\begin{array}{c}
\left\langle \psi \left( x\right) \mu \left( x_{0}\right) \sigma \left(
Px_{0}\right) \right\rangle _{0} \\ 
\left\langle \bar{\psi}\left( x\right) \mu \left( x_{0}\right) \sigma \left(
Px_{0}\right) \right\rangle _{0}%
\end{array}%
\right) - \\[0.05in]
-i\frac{m}{2}\partial \left( 
\begin{array}{c}
\left\langle \psi \left( x\right) \mu \left( x_{0}\right) \sigma \left(
Px_{0}\right) \right\rangle _{0} \\ 
\left\langle \bar{\psi}\left( x\right) \mu \left( x_{0}\right) \sigma \left(
Px_{0}\right) \right\rangle _{0}%
\end{array}%
\right) -i\frac{m}{2}\left( 1-2\lambda \right) \bar{\partial}\left( 
\begin{array}{c}
\left\langle \psi \left( x\right) \mu \left( x_{0}\right) \sigma \left(
Px_{0}\right) \right\rangle _{0} \\ 
\left\langle \bar{\psi}\left( x\right) \mu \left( x_{0}\right) \sigma \left(
Px_{0}\right) \right\rangle _{0}%
\end{array}%
\right) + \\[0.05in]
+\bar{\partial}^{2}\left( 
\begin{array}{c}
\left\langle \psi \left( x\right) \mu \left( x_{0}\right) \sigma \left(
Px_{0}\right) \right\rangle _{0} \\ 
\left\langle \bar{\psi}\left( x\right) \mu \left( x_{0}\right) \sigma \left(
Px_{0}\right) \right\rangle _{0}%
\end{array}%
\right)
\end{multline}%
Substituting it in (\ref{2-5}) and evaluating the left-hand side we obtain
the following differential equation for local magnetization $\bar{\sigma}%
\left( t\right) $:%
\begin{multline}
\left( G+\tilde{G}\right) \bar{\sigma}^{\prime \prime }-\left[ G^{\prime }+%
\tilde{G}^{\prime }-G+\lambda \left( G+\tilde{G}\right) \right] \bar{\sigma}%
^{\prime }+ \\[0.05in]
+\frac{1}{4}\left[ G^{\prime \prime }+\tilde{G}^{\prime \prime }-\frac{1}{t}%
\left( G^{\prime }+\tilde{G}^{\prime }\right) -2G^{\prime }-\tilde{G}%
+2\lambda \left( G^{\prime }+\tilde{G}^{\prime }+\tilde{G}\right) \right] 
\bar{\sigma}=0
\end{multline}%
It can be brought to a simpler form (\ref{16}) by means of substitution (\ref%
{15}).

\section{Acknowledgements}

I am especially grateful to Y. P. Pugai for his interest to my work and
encouragement. I am also grateful to my scientific advisers A. I. Bugrij and
V. N. Shadura for giving me creative freedom and to N. Z. Iorgov and V. N.
Shadura for pointing out the paper \cite{SchE}.

\end{document}